\documentclass[preprint,aps]{revtex4}

\usepackage{graphicx}
\usepackage{dcolumn}
\usepackage{bm}

\begin{document}

\preprint{}
\pacs{07.79.Cz, 74.55.+v, 84.37.+q}
\title{A Scanning Tunneling Microscope for a Dilution Refrigerator}

\author{M.\ Marz}
\email{Michael.Marz@kit.edu}
\affiliation{Physikalisches Institut, Karlsruher Institut f\"ur Technologie, 76131 Karlsruhe}
\author{G.\ Goll}
\email{Gernot.Goll@kit.edu}
\affiliation{Physikalisches Institut, Karlsruher Institut f\"ur Technologie, 76131 Karlsruhe}
\author{H.\ v.\ L\"{o}hneysen}
\affiliation{Physikalisches Institut and Institut f\"ur Festk\"orperphysik,\\ Karlsruher Institut f\"ur Technologie, 76131 Karlsruhe}
\date{\today}

\begin{abstract}
We present the main features of a home-built scanning tunneling microscope that has been attached to the mixing chamber of a dilution refrigerator. It allows scanning tunneling microscopy and spectroscopy measurements down to the base temperature of the cryostat, $T\approx30\,$mK, and in applied magnetic fields up to $13\,$T. The topography of both highly-ordered pyrolytic graphite (HOPG) and the dichalcogenide superconductor NbSe$_2$ have been imaged with atomic resolution down to $T\approx50\,$mK as determined from a resistance thermometer adjacent to the sample. As a test for a successful operation in magnetic fields, the flux-line lattice of superconducting NbSe$_2$ in low magnetic fields has been studied. The lattice constant of the Abrikosov lattice shows the expected field dependence $\propto 1/\sqrt{B}$ and measurements in the STS mode clearly show the superconductive density of states with Andreev bound states in the vortex core.
\end{abstract}
\maketitle

\section{Introduction}
Thirty years after the invention of the scanning tunneling microscope (STM) by Binnig and Rohrer\,\cite{1} the interest in this technique is still increasing. STM and scanning tunneling spectroscopy (STS) measurements on superconductors are one example of this emerging field. High-temperature superconductors have been successfully investigated by STM and STS\,\cite{2,3}. For the study of heavy-fermion superconductors with transitions temperatures of about 1\,K and below, STM and STS measurements at very low temperatures are required. A STM operating at such low temperatures would make it feasible to investigate properties of unconventional superconductors, e.\ g., the order parameter in UPt$_3$. To observe the full opening of the superconducting gap, the temperature has to be sufficiently below the critical temperature, e.\ g., $T_c\approx 550\,$mK for UPt$_3$\,\cite{4}.\\
In the last years, STMs for low temperatures down to $300\,$mK and moderate magnetic fields have become commercially available. However, for even lower temperatures and high magnetic fields $B\ge 10\,$T only a small number of home-built scan heads exist\,\cite{5}. The implementation of an STM into a dilution refrigerator is not straightforward, there are several requirements in order to successfully run the system, i.\ e., mechanical and electrical noise damping, a functional coarse approach for low temperatures, etc.\\
In this article we describe a home-built STM for low temperatures $T\approx 30\,$mK that works in a standard dilution refrigerator and in magnetic field up to $B=13\,$T. Calibration measurements on highly-ordered pyrolytic graphite (HOPG) and NbSe$_2$ at room temperature are presented. Calibration at low temperature and in magnetic fields has been performed on NbSe$_2$ in the Shubnikov phase. The measurements demonstrate the good spatial and energy resolution, respectively a high voltage resolution of the order of $1\,$K and stability of the instrument.
\section{Setup}
The low-temperature STM has to meet several stringent requirements. First of all the setup has to be rigid enough to provide mechanical stability and a good thermal contact to the mixing chamber. The choice of appropriate materials for temperature- and field-dependent measurements should take into account the thermal expansion and the magnetic properties. The setup should be electronically isolated from ground loops. Electromagnetic stray fields should be shielded and filtered from the leads.\\
In this section the general setup is discussed briefly. The STM is based on the design developed for scan heads used at higher temperatures\,\cite{6,7} but with a modified tip approach in order to allow for measurements in magnetic fields. The most important features of the design are its versatility and compactness which allows to fit it into the cryostat tail within a superconducting solenoid. The setup allows cycle times of about two to three days combined with a high mechanical stability. During refill with liquid helium, mechanical distortions occure, nevertheless the tip does not need to be retracted completely from its position close to the sample surface. This advantage compensates for the long warming and cooling time necessary for tip and sample changes.
\begin{figure}[ht]
\includegraphics[width=0.5\linewidth]{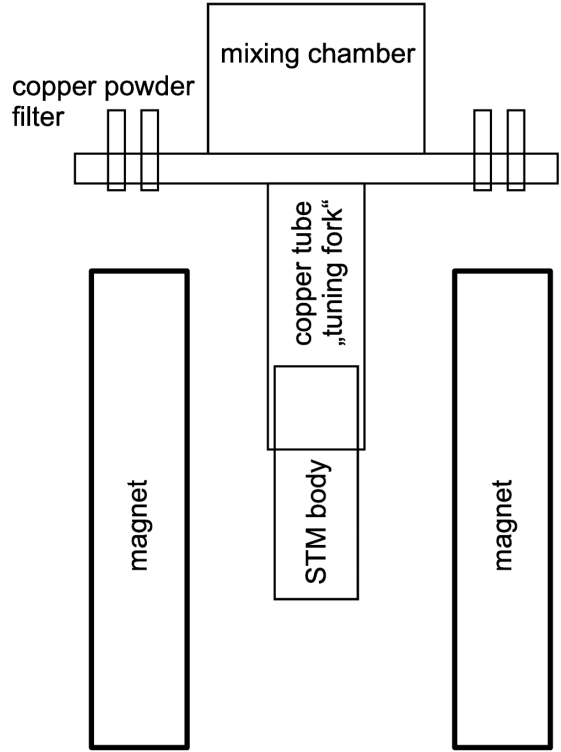}
\caption{\label{fig:skizze} Sketch of the principal setup mounted to the mixing chamber. The illustration is not to scale. A copper plate supporting the copper-powder filters is mounted directly to the mixing chamber. An incised copper tube is used to ensure that the STM body, more precisely the sample position, is located in the homogeneous part of the magnetic field generated by a superconducting solenoid.}
\end{figure}
\subsection{General Assembly}
A standard $^3$He/$^4$He dilution refrigerator (GVL Leiden Cryogenics, Leiden, Netherlands) with a continuous 1\,K plate and a base temperature well below $T_{\rm{min}}\approx30\,$mK is used to host the home-built STM. The cryostat is equipped with a superconducting solenoid installed outside the vacuum chamber and designed to reach magnetic fields up to $B_{\rm{max}}=13\,$T. The general setup is sketched in figure~\ref{fig:skizze}. The STM body is made of highly resistive stainless steel (fig.~\ref{fig:bauteile} A) in order to reduce eddy currents, when high magnetic fields are applied. Furthermore it is supposed to be nonmagnetic, although a small magneto mechanical effect, e.\ g., due to magnetostriction can not be excluded, when high magnetic fields are applied.\\
The STM sample position is placed in the center of the magnet bore for having a homogeneous field. A heater made of a twisted-pair manganin wire allows heating of the whole scan head. It is wound around the stainless steel body and thermally anchored but electrically isolated with stycast epoxy (Stycast  2850FT Polymer). To avoid local hot spots, the heater is glued in a meander-like fashion to cover most of the surface. The temperature of the STM body is controlled by a Cernox bare chip thermometer calibrated in the temperature range from $T=30\,$mK to $T=1.5\,$K. The sensor is mounted inside an incised  Cu screw into the STM body to allow good thermal contact between thermometer and STM body especially at low temperatures. The STM body itself is thermally anchored to the mixing chamber via an oxygen-free copper tube. In addition, a heat shield thermal anchored to the mixing chamber surrounds the STM body. It has been verified that a temperature gradient of less than $20\,$mK occurs between mixing chamber and STM body. As a test we recorded tunnel spectra of superconducting aluminum ($T_c=1.1\,$K) which exhibited a temperature dependence down to a STM body temperature of $\approx 100\,$mK. We observed a temperature dependent $dI/dV$ curve down to $200\,$mK, i.e., $T/T_c=0.18$. The fact that we did not observe a further sharpness of the BCS singularity at lower temperature is probably due to electronic noise. Thus we can conclude in this particular experiment (STM body temperature $\approx 100\,$mK) that the sample temperature is at least $200\,$mK. All temperatures given in this article correspond to the STM body temperature determined in the experiment from the calibrated sensor.
\subsection{STM}
\begin{figure*}[ht]
\includegraphics[angle=90,width=\linewidth]{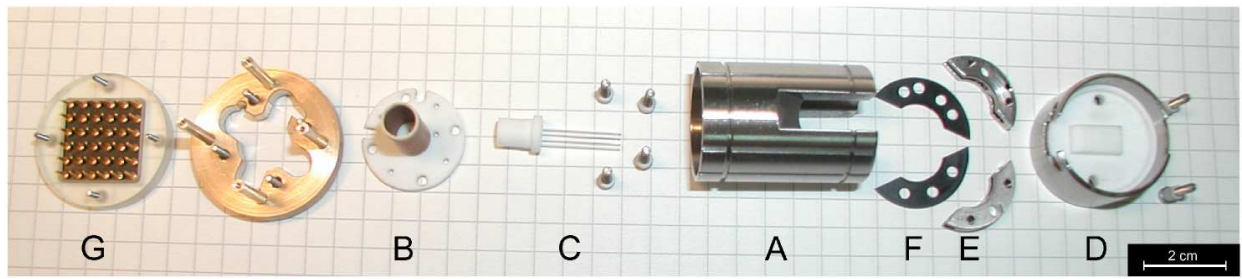}
\caption{\label{fig:bauteile} Photograph of the single parts of the scan head before assembly. (A) stainless steel body, (B) piezo tube, (C) sapphire roads, (D+E) sample holder (F) isolation and (G) plug for electrical connection of the piezo scan head, thermometers and heater.}
\end{figure*}
\begin{figure}[ht]
\includegraphics[width=0.1\linewidth]{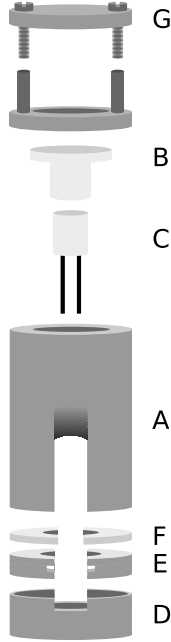}
\caption{\label{fig:assembly} Sketch of the assembling of the parts from fig~\ref{fig:bauteile}. All parts are fixed with insulating screws, except for the piezo (B) and the sapphire roads (C) which are glued with Torr Seal and the 36-pin plug (mounted inside G) which is glued with stycast epoxy (Stycast  1266 Polymer).}
\end{figure}

The STM consists of a scan-head unit with a commercial five-electrode piezo tube scanner (EBL \#2, Staveley NDT Technologies, East Hartford, USA) (fig.~\ref{fig:bauteile} B). The inner electrode of the piezo tube is used for the $z$ movement and the outer electrode is split into four parts, for $\pm\,x$ and $\pm\,y$ movement, respectively. The way of assembling the scan head from the parts of fig.~\ref{fig:bauteile} is displayed in fig.~\ref{fig:assembly}. Almost all parts are fixed with metallic screws to ensure good thermal coupling, only the piezo is glued with Torr Seal epoxy (Varian) to a macor\,\cite{macor} panel and the sapphire rods.\\
The coarse approach of the STM tip to the sample is done by a vertical slip-stick mechanism using the $z$ movement. The tip is attached to a non-magnetic CuBe slider which is supported by two sapphire rods (fig.~\ref{fig:bauteile} C).
The coarse approach of the tip, or rather the coarse approach of the slider, is based on inertia movement introduced by Pohl\,\cite{8}. 
\begin{figure}[ht]
\includegraphics[width=0.4\linewidth]{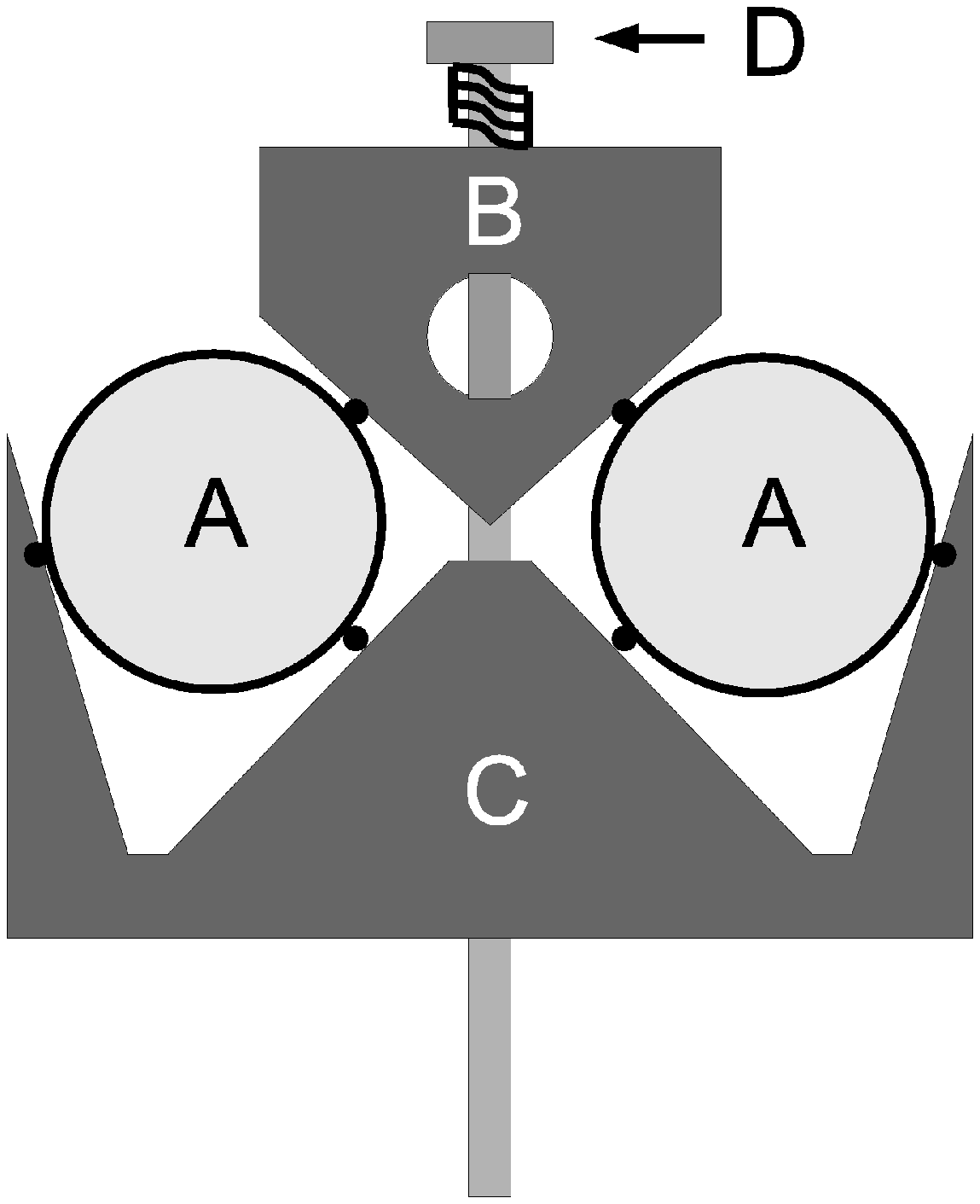}
\caption{\label{fig:schlitten} Sketch of the cross-section slider (parts B,C,D) mounted on the sapphire rods (A). The V/W shape leads to three lines at which the slider is touching each rod. A displacement of the sapphire rods due to horizontal forces is avoided by the V/W shape of the slider.}
\end{figure}
The cross-section of the slider is sketched in figure~\ref{fig:schlitten}. The upper part (fig.~\ref{fig:schlitten} B) consists of a CuBe cube with a cylindrical hole and three perpendicular drilling holes. One hole is used to fix the tip, another one is used for the connections of the current lead. The lower part (fig.~\ref{fig:schlitten} C) is connected through the middle hole to the upper part by a screw (fig.~\ref{fig:schlitten} D) and pre-loaded by a spring which allows to apply a variable pressure to the sapphire rods (fig.~\ref{fig:schlitten} A). This changes the friction between the slider and the sapphire rods and therefore permits the control of the speed, in particular the step width, of the coarse approach. The step width has to be large enough at room temperature to maintain a sufficient displacement at low temperatures. The V/W shape of the slider prevents a displacement of the sapphire rod due to horizontal forces.\\
The STM tips are prepared by cutting a $250\,\mathrm{\mu}$m Pt/Ir (80/20) wire in tension with a ceramic scissors under an angle of $45^\circ$. The typical length of the tip is about $2-3\,$mm. The tip is fixed with a CuBe screw to the slider, allowing a good thermal contact and high mechanical stability.\\
The sample is glued with silver conductive paint to a copper holder, which is fixed with two isolated bolts at the bottom of the main body (fig.~\ref{fig:bauteile} D). The copper holder with the sample is supported by stainless steel shoes (fig.~\ref{fig:bauteile} E) that are also isolated from the main body (fig.~\ref{fig:bauteile} F). To get rid of possible contamination, e.\ g., a water film on the sample surface, after a change of sample and tip at ambient conditions, the scan head is heated to $100^{\circ}\,$C, while the inner vacuum can is pumped. The outer parts, i.\ e., the magnet and the He dewar are cooled to liquid nitrogen temperature meanwhile. Thus, we can expect that nearly all water evaporates from the sample surface and condenses at the wall of the vacuum chamber. To control the temperature during heating a Pt-100 thermometer is attached on the reverse side of the sample holder (fig.~\ref{fig:bauteile} D). No further preparation is installed at present. However, sample space in the vacuum chamber allows for the installation of an in-situ cleavage station.\\
The electrical wiring for the piezo scan head, the thermometers and the heater is done via a 36-pin IC socket. The leads from the plug to the mixing chamber (including the copper-powder filters, see below) are made of polyimid-isolated $100-\mathrm{\mu}$m Cu wire. For the connection to the room temperature part of the cryostat manganin wires are used for the piezo leads to minimize the thermal conductance and to avoid a modification of the signal form due to large resistance changes. The thermometer leads from the mixing chamber to the 1\,K pot are made of superconducting NbTi wires.\\
The feedthroughs for current and voltage are made of gold-plated copper-shielded coaxial cable with stranded copper center wire (LakeShore CC-SC-100), which is connected via SMA plugs to the copper-powder filter placed at the mixing chamber. The connection to the upper part of the cryostat is done with stainless steel coaxial cables (LakeShore CC-SS-100).\\
The measurement of the capacitance between tip and sample is used to monitor and control the tip-sample distance during cooling the cryostat and at low temperatures. The maximum scanning area of the setup at room temperature is about $10\, \mathrm{\mu}$m$\times 10\, \mathrm{\mu}$m, at low temperatures it is reduced to $2\, \mathrm{\mu}$m$\times 2\, \mathrm{\mu}$m. 

\subsection{Electronics}
The feedback loop, the high voltage amplification, and data acquisition electronics are commercially available units (ECS, Cambridge, UK). For STS measurements the output voltage is modulated with a frequency generator (Stanford Research Systems DS345, Sunnyvale). The current signal is deconvoluted with an analog lock-in amplifier (Ithaco 3916B, Ithaca, NY). For the current amplification, a current-voltage transformator with variable gain (Femto DHCL 200, Berlin) is used. The data acquisition is performed with a standard PC equipped with a digital signal processing (DSP) card.
\subsection{Electrical and Mechanical Filtering}
\begin{figure}[ht]
\includegraphics[angle=90,width=0.7\linewidth]{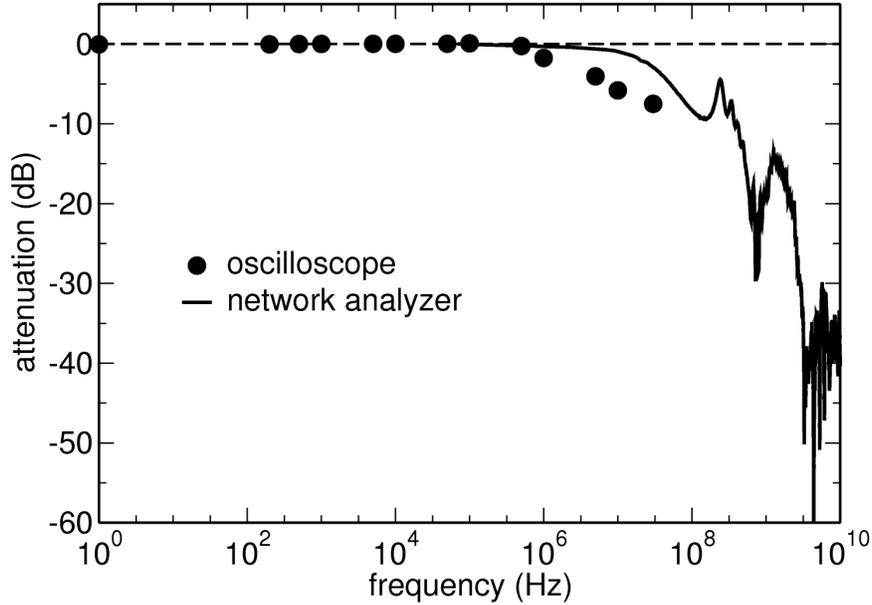}
\caption{\label{fig:daempfung} Damping behavior of the copper-powder filter. The measurement of the solid line was performed with a network analyzer in a frequency range $70\,$kHz$-20\,$GHz. The low-frequency data are determined by applying an ac signal on one end of the filter and measuring the response signal on the other end with an oscilloscope.}
\end{figure}

Electrical and mechanical filtering is the most important challenge for proper STM and STS performance. Main reasons for the electrical noise are ground loops and long leads, mechanical noise occurs mainly through vibrations of the pumping lines (in our case mostly of the $1\,$K pumping line and, to a lesser extant, the still pumping line) and through boiling of the cryogenic fluids. To reduce vibrational noise no liquid nitrogen shield was used, but super insulation foil in the outer vacuum can instead. Additionally, the He dewar was pressurized to $50-100\,$mbar above athmospheric pressure to reduce boil-off. By pumping on the He dewar no further advantage was gained.\\
To minimize electrical noise, the STM electronics is electrically isolated from all other electronics as well as from all pumping lines. This permits the use of the cryostat as the only ground in the whole system. All leads used for measurements and control are filtered by home-built copper-powder filters (skin-effect filters). These filters are mounted on top of the STM holder next to the mixing chamber. Their main purpose is to reduce high-frequency noise. The attenuation vs. frequency curve is shown in fig~\ref{fig:daempfung}. 
A further advantage of these filters is the good thermal coupling of the leads to the mixing chamber. The filters consist of $2\,$m of copper wire, copper powder (grain size $50\,\mathrm{\mu}$m) and a copper jacket that is attached very closely to the mixing chamber and therefore cools down the electronic system. Low-pass filters (cutoff frequency  $\nu=2\,$kHz) are installed behind the high-voltage amplifier to suppress noise of the used electronics. An additional low-pass filter with a cutoff at $4\,$kHz is installed consecutively to the preamplifier.\\
The cryostat itself is hanging from a platform supported by three air cushions for damping the mechanical noise. Typical cut-off frequency of the low-pass system is $10\,$Hz. For additional damping of mechanical distortions (vibrations of pumping lines etc.) the scan head unit is fixed in an incised copper tube. The tube acts like a tuning fork, with a sharp resonance and damping of all other frequencies.
\section{Calibration at Room Temperature}
After  assembling the instrument, first measurements were carried out at room temperature and at ambient pressure to calibrate the scanning unit. First, the calibration for large scan areas are performed on a commercially available calibration device ('NanoGrid', Schaefer Technologie GmbH, Langen). The grid is a $2 \,$mm$\,\times\,2\,$mm sized polymer chip with crossed lines at a relative distance of $d=160\,$nm. In order to make the surface conductive, AuPd is evaporation-deposited on top. From these measurements scaling factors of $30.0\,$nm/V for $\pm\,$x movement and $43.1\,$nm/V for $\pm\,$y movement are obtained.\\
\begin{figure}[ht]
\includegraphics[width=.7\linewidth]{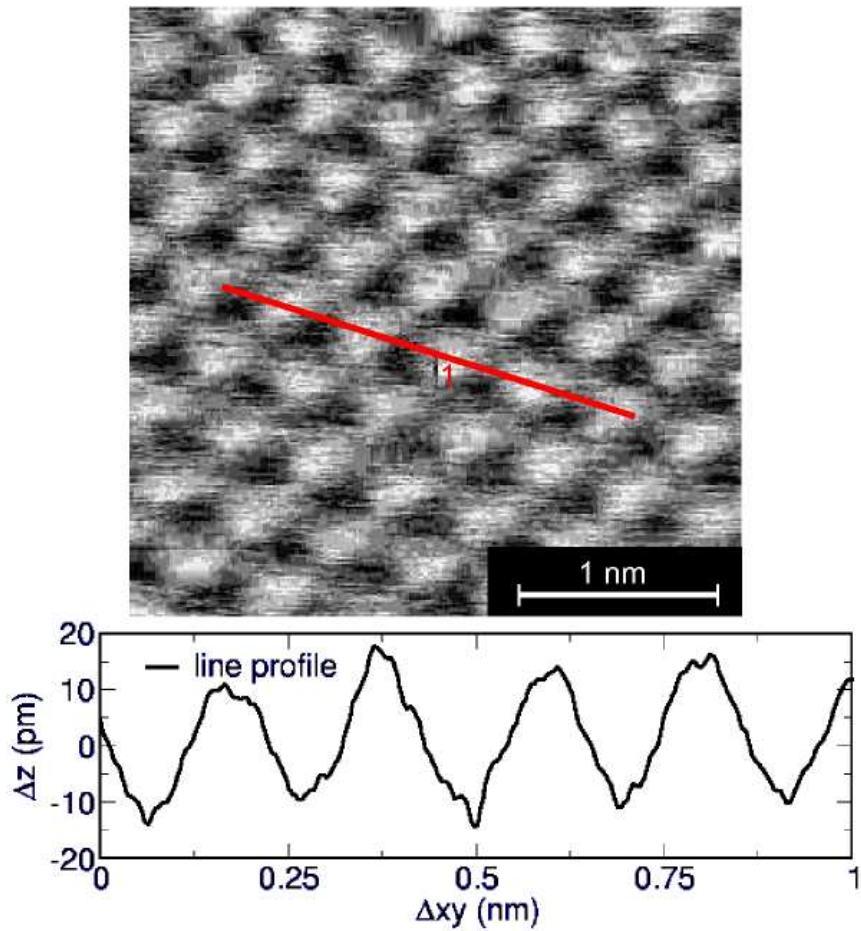}
\caption{\label{fig:hopg} Raw data of the atomic resolution on HOPG ($V_{bias}=76\,$mV, $I= 2\,$nA) at room temperature and ambient pressure. The lower graph shows the line profile marked in the scan. The well-known lattice constant was used to calibrate the scan head for room temperature and small scanning areas.}
\end{figure}
HOPG is used for the calibration of the piezo positioning elements, as is often done as a standard procedure for two reasons: First, it is relatively easy to get atomic resolution on HOPG, and second, the preparation of a clean surface is very simple. As HOPG has a layered structure, a fresh surface can be obtained by using adhesive tape to remove some of the top layers. The lattice parameters are well known from literature: the nearest neighbor distance is $0.142 \,$nm, whereas the in-plane lattice constant is  $0.246 \,$nm\,\cite{9}. To achieve atomic resolution a higher spatial sensitivity for the scanning unit is needed. To increase the resolution at room temperature a voltage divider could be used optionally. The applied voltage to the scan piezo is divided by a factor of 10 to avoid the digitization artifacts. These steps appear in the control voltage of the piezo amplifier, which has a limited resolution of $12$ bit. Figure~\ref{fig:hopg} shows the raw data of the hexagonal atomic structure of the HOPG sample at room temperature at ambient conditions.
\section{Calibration at Low Temperature and in Magnetic Field}
\begin{figure}[ht]
\includegraphics[width=.5\linewidth]{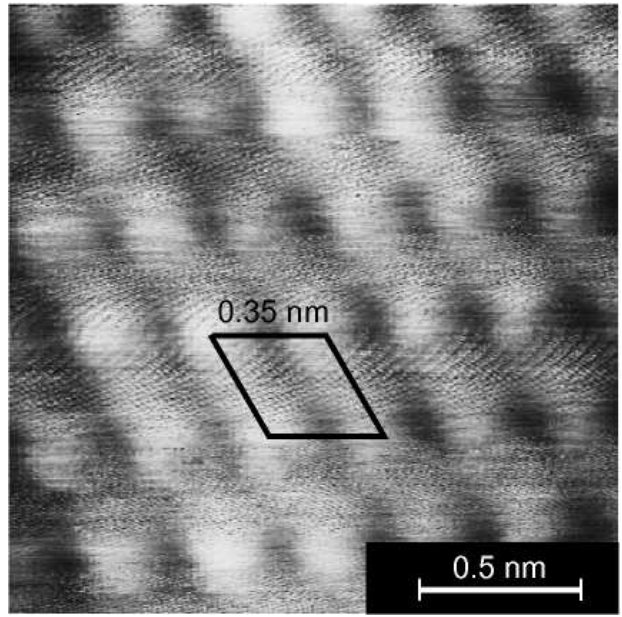}
\caption{\label{fig:nbse_2_tt-mark} Atomic resolution on NbSe$_2$ at $T \approx 50\,$mK, with $V=322.3\,$mV and $I=1.7\,$nA. The rombus indicates the unit cell of the hexagonal lattice. With this data the scan head was calibrated for low temperature and small scanning areas.}
\end{figure}
\begin{figure}[ht]
\includegraphics[angle=90,width=.7\linewidth,clip=]{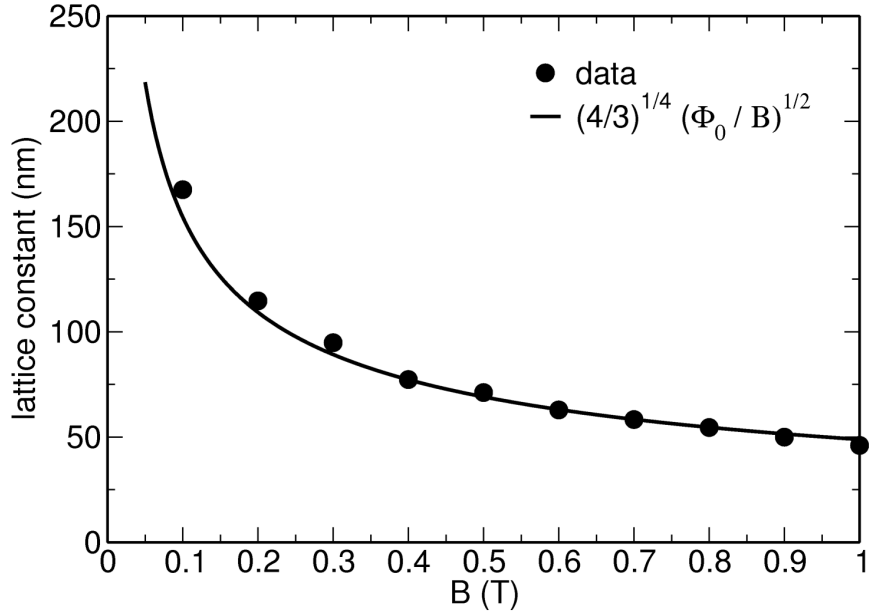}
\caption{\label{fig:gitter} Field dependence of the flux line lattice constant. The curve is calculated without any free parameter. The expected $a\propto (1/B)^{1/2}$ dependence is clearly seen in the measurements.}
\end{figure}

Besides HOPG, the layered type-II superconductor NbSe$_2$ with a critical temperature of $7.2\,$K was investigated. This material is a favorite object for STM and STS studies\,\cite{10,11,12,13} because it has a layered structure similar to HOPG and the sample preparation can be done in a similar way. Figure~\ref{fig:nbse_2_tt-mark} displays a topographic scan on NbSe$_2$ taken at $T \approx 50\,$mK. The in-plane lattice constant is $0.345\,$nm\,\cite{14}. From the comparison to the lattice constant obtained at room temperature and at low temperature the conversion factor between the room-temperature and the low-temperature calibration of the scan head can be calculated. This conversion factor is determined to be $\approx 0.2$ for small scanning areas.\\
\begin{figure*}[ht]
\includegraphics[angle=90,width=\linewidth]{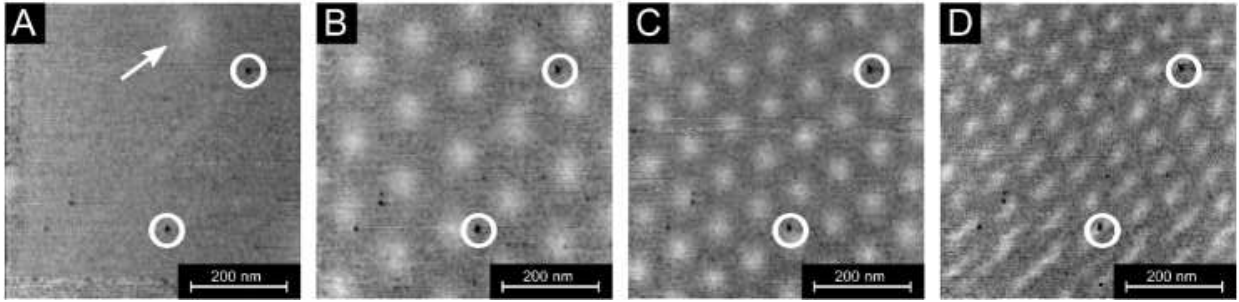}
\caption{\label{fig:0T-flussliniengitter} Measurements of the vortex lattice in NbSe$_2$, with $V=1.6\,$meV and $I=2\,$nA at $T\approx 100\,$mK for different fields $B=0$ (A), $B=0.2\,$T (B), $B=0.4\,$T (C) and $B=0.8\,$T (D). The bright spot in figure~A (arrow) shows a trapped flux line. The circles mark surface defects. They do not change the position with increasing field indicating that no major drift is observable. 
}
\end{figure*}
In magnetic fields above the lower critical field, type-II superconductors like NbSe$_2$ enter the Shubnikov phase and magnetic flux can penetrate the sample in the form of quantized flux lines, regions where the superfluid density vanishes, i.\ e., regions with a normal conducting core. These flux lines are regularly arranged (Abrikosov lattice) in superconductors with low pinning and the change of the electronic structure between the superconductive to the normal-conductive density of states is directly observable with STM measurements. To observe the flux-line lattice of NbSe$_2$ a small magnetic field of $\le 1\,$ T has been applied. Field-dependent measurements of the local density of states in the constant current mode were performed at applied voltage $V\approx 1.6\,$mV. At this voltage the maximum of the coherence peak is observed in the tunneling spectra in zero field (see fig.~\ref{fig:spektroskopie}) and the maximal contrast between normal and superconducting regions is obtained.\\
In figure~\ref{fig:0T-flussliniengitter}, four STM images of the same surface region acquired in the constant current mode at $V\approx 1.6\,$mV and $T\approx 100\,$mK are shown for different applied magnetic fields. The image (A) is measured at zero field, but after a field sweep. A trapped flux line in the upper right-hand corner can be identified (arrow). Flux can still be present in the sample caused  either by a remanent field of the superconducting magnet or by pinning of a flux line at a defect. Image (B) is taken at a magnetic field of $B=200\,$mT, the expected hexagonal structure of the Abrikosov lattice is clearly observed. Raising the magnetic field (C and D) further increases the density of vortices. For larger fields $B\ge 800\,$mT a noticeable drift at the lower parts of the scan area occurs (fig.~\ref{fig:0T-flussliniengitter} D), which possibly is due to heating of the setup by eddy currents while changing the field. However major shifts due to heating of the sample and/or the tip during field change can be excluded. As seen from the pictures (A-D) of figure~\ref{fig:0T-flussliniengitter}, the overall drift of the scanning area between different scans can be neglected, since characteristic surface defects remain at almost the same positions in every scan (white circles).\\
In order to demonstrate the performance of the STM in magnetic fields the lattice constant $a$ of the Abrikosov lattice has been extracted from the data. In figure~\ref{fig:gitter} the lattice constant $a$ is shown together with the expected field dependence\,\cite{15} $a=\left(4/3\right)^{1/4}\cdot\left(\Phi_0/B\right)^{1/2}$ where $\Phi_0=h/2e$, without any adjustable parameters. Therefore, imaging the vortex lattice allows a calibration of our scan head at low temperatures for large scanning areas, in addition to the calibration for small scanning ranges achieved by atomic resolution. The scaling factor with respect to the room-temperature motion of the scan head is $\approx 0.2$ in both cases.\\
\section{Spectroscopy at Low Temperatures}
\begin{figure}[ht]
\includegraphics[angle=90,width=.7\linewidth,clip=]{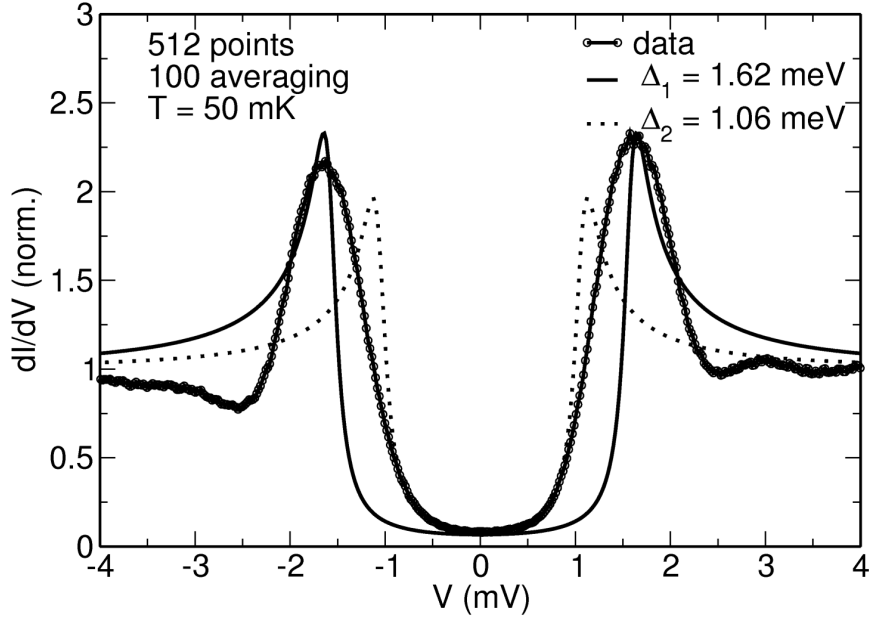}
\caption{\label{fig:spektroskopie} Differential conductance of a NbSe$_2$-PtIr tunnel contact at $T\approx 50\,$mK. The data are taken at one point on the surface averaged over $100$ measurements with $512$ steps each in this voltage span. The solid and dashed lines are calculated using the BTK theory, details are given in the text. The deviations between the fit are probably due to  charge-density wave dynamics or to the gap anisotropy in NbSe$_2$.}
\end{figure}
While running the scan head in the STM mode gives insights into the topography of the sample surface, the STS mode directly yields the electronic density of states at the surface via the differential conductance vs. voltage curves. The characteristics of the superconducting density of states in NbSe$_2$ are expected to occur at $\Delta/e=1.75\cdot k_{\rm{B}}T_c/e=1.08\,$mV, i.\ e., at low DC bias. To resolve signatures at such low energies a high energy resolution, respectively voltage resolution $ \delta V < 1\,$meV is needed. For that purpose a switchable voltage divider (1:1, 1:10, 1:100, 1:1000) was installed to increase the resolution of the electronics. In order to record the tunneling curves with high resolution the tunneling voltage is modulated with an AC bias ($V_{AC}\approx 20\,\mathrm{\mu}$V) and the current signal is measured with a lock-in amplifier.\\ Figure~\ref{fig:spektroskopie} shows a typical measurement of the differential conductance of a NbSe$_2$-PtIr tunnel contact (circles) at $T=50\,$mK in zero field. At such low temperatures the spectrum shows pronounced conductance peaks due to the BCS singularity of the density of states\,\cite{15a}. The value of the energy gap is determined by fitting the spectra using the BTK theory\,\cite{16} for a large barrier parameter (tunnel limit). By fitting the positions at the maxima (solid line), the energy gap $\Delta=1.62\,$meV is obtained, whereas a fit to the width of zero conductance (dashed line) results in $\Delta=1.06\,$meV for the gap. These variations reflect a notable gap anisotropy in this material\,\cite{13} most probably caused by the presence of charge-density wave dynamics in NbSe$_2$ since our measurements are relatively slow.  On the other hand a dynamic broadening due to charge density waves is also possible.\\
\begin{figure}[ht]
\includegraphics[width=.907\linewidth]{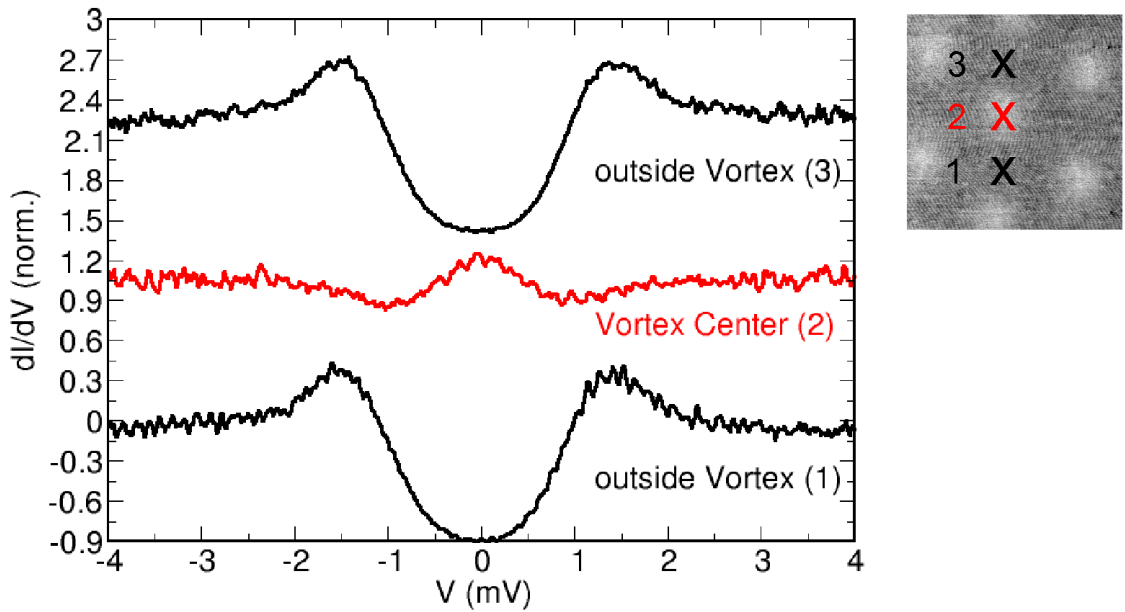}
\caption{\label{fig:boundstate} Differential conductance of a NbSe$_2$-PtIr tunnel contact in the vortex state (left panel) at $T\approx 100\,$mK. The positions of the PtIr tip above the NbSe$_2$ sample are marked (right panel). Spectra 1 and 3 are measured in the superconducting regions outside the vortex, spectrum 2 is measured directly at the center of the vortex. Spectra are shifted along the y axis for clarity.}
\end{figure}
In an applied field $B=500\,$mT, line scans ('topographic mode') through a vortex were performed (not shown) in addition to the images of the vortex lattice. The spectra recorded along the line scan clearly show a somewhat broadened BCS-like density of states in the superconducting regions and an enhancement of the differential conductance at zero bias in the normal-conducting regions of the vortex center which can be attributed to Andreev bound states\,\cite{17}. This enhancement in the conductance has been reported earlier\,\cite{18,19} and is very sensitive to the position of the tip in respect to the vortex center.\\
As representative examples, three different spectra are shown in figure~\ref{fig:boundstate}. The data acquired at a particular point on the surface were averaged over $100$ measurements in the same voltage range of $|V|\le4\,$mV with $512$ steps each in this voltage span. Measurements (1) and (3) were performed away from a vortex and show the superconductive density of states. The spectrum at the vortex (2) clearly shows the enhancement at zero bias of the conductance due to the Andreev bound states. The observation of this state indicates the stable and reproducible positioning of the tip at the center of the vortex.
\section{Acknowledgments}
We thank R. Hoffmann, B. Pilawa, C. Debuschewitz and W. Wulfhekel for helpful discussions. Many thanks to C. P\'erez Le\'on for the help with the assembly of the electronics. We thank D. Rosenmann at the Argonne Nation Laboratory, Condensed Matter Division, for providing us the NbSe$_2$ sample.


\begin{thebibliography}{}
\bibitem{15a} An enhancement up to a factor of 2.2 of the conductance peaks with respect to the high-bias value was observed. This value is of the same magnitude as literature values for NbSe$_2$ at $T=100\,$mK\,\cite{20}, which confirm that an electronic temperature less than $100\,$mK had been obtained.
\end{thebibliography}
\end{document}